\documentclass[fdp,fleqn]{w-art}
\usepackage{times}
\usepackage{w-thm}
\usepackage{amsmath,amsfonts,amsbsy,amssymb}
\usepackage{indentfirst}
\usepackage{mathtools}
\usepackage[]{graphicx}

\def\be{\begin{equation}}
\def\ee{\end{equation}}
\def\bse{\begin{subequations}}
\def\ese{\end{subequations}}
\def\bal{\begin{align}}
\def\ealn{\end{align}}

\begin{document}
%%    The information for the title page will be placed between
%%    \begin{document} and \maketitle. The order of most entries
%%    is determined by the class file and can not be changed by
%%    rearranging them. The maketitle command follows after the
%%    abstract.
%%
%%    Most of the following commands will be completed by the publisher.
%%
%%    The copyrightyear is defined in the .clo file as the first argument
%%    of the copyrightinfo command. If the copyrightyear differs from that
%%    value it might be adjusted by the following definition:
%%
%% \renewcommand{\copyrightyear}{2003}% uncomment to change the copyrightyear.
%%
\DOIsuffix{theDOIsuffix}
%%
%% issueinfo for header and copyright line
\Volume{55}
\Issue{1}
\Month{01}
\Year{2007}
%%
%%    First and last pagenumber of the article. If the option
%%    'autolastpage' is set (default) the second argument may be left empty.
\pagespan{1}{}
%%
%%    Dates will be filled in by the publisher. The 'reviseddate' and
%%    'dateposted' (Published online) entry may be left empty.
\Receiveddate{Date Month 2014}
\Reviseddate{Date Month 2014}
\Accepteddate{Date Month 2014}
\Dateposted{Date Month 2014}
\keywords{Quantum Nambu geometry, topological insulator, quantum Hall effect, monopole}

%% \pretitle{Editor's Choice}

%% We have a short and a long form for the title. The short form
%% (optional argument) goes into the running head.

\title[NCG in higher D. QHE in A-class TI]{Non-Commutative Geometry in Higher Dimensional Quantum Hall Effect as  A-Class  Topological Insulator} 

%% Please do not enter footnotes or \inst{}-notes into the optional
%% argument of the author command. The optional argument will go into
%% the header.  If there is only one address the marker \inst{x} may be
%% omitted.

%% Information for the first author.
\author[K. Hasebe]{Kazuki Hasebe\inst{}%
  \footnote{Corresponding author\quad E-mail:~\textsf{hasebe@dg.kagawa-nct.ac.jp}, 
            Phone: +81\,875\,83\,8528, 
            Fax: +81\,875\,83\,6389}}
\address[\inst{1}]{Kagawa National College of Technology, Takuma-cho, Mitoyo, Kagawa 769-1192, Japan}
%%
%%    Information for the second author
%\author[S. Author]{Second Author\inst{1,2,}\footnote{Second author footnote.}}
%\address[\inst{2}]{Second address}
%%
%%    Information for the third author
%\author[T. Author]{Third Author\inst{2,}\footnote{Third author footnote.}}
%%
%%    \dedicatory{This is a dedicatory.}
\begin{abstract}
 
 We clarify relations between the higher dimensional quantum Hall effect and A-class topological insulator. In particular, we elucidate  physical implications of the higher dimensional non-commutative geometry in the context of A-class topological insulator. 
 This presentation is based on \cite{Hasebe2014-1}. 

\end{abstract}
%% maketitle must follow the abstract.
\maketitle                   % Produces the title.

%% If there is not enough space inside the running head
%% for all authors including the title you may provide
%% the leftmark in one of the following three forms:

%% \renewcommand{\leftmark}
%% {F. Author: A short title}

%% \renewcommand{\leftmark}
%% {F. Author and S. Author: A short title}

%% \renewcommand{\leftmark}
%% {F. Author et al.: A short title}

%% \tableofcontents  % Produces the table of contents.
%%%%%%%%%%%%%%%%%%%%%%%%%%%%%%%%%%%%%%%%%%%%%%%%%%%%%%%%%%%%%%%%%%%%%%%%%%%%%%%%%%%%%%%%%%%%
\section{Introduction}
%%%%%%%%%%%%%%%%%%%%%%%%%%%%%%%%%%%%%%%%%%%%%%%%%%%%%%%%%%%%%%%%%%%%%%%%%%%%%%%%%%%%%%%%%%%%%
%\subsection{First subsection}

 The non-commutative geometry (NCG) is the underlying geometry of the quantum Hall effect (QHE).  In the lowest Landau level (LLL), the electron coordinates are identified with the center of mass coordinates that satisfy the NC algebra: 
%%%%%%%%%%%%%%%%%%
\begin{equation}
[X, Y]=i{\ell}^2. 
\end{equation}
%%%%%%%%%%%%%%%%%%
Almost a decade ago, the time-reversal symmetric counterpart of QHE, dubbed as quantum spin Hall effect, was theoretically proposed \cite{KaneMele2005} and experimentally confirmed \cite{Koenigetal2007}.  Subsequently, the 3D version of the quantum spin Hall effect, topological insulator (TI), was also discovered 
\cite{FuKane2007}. Now we understand there exist a variety of topological classes of QHE with different symmetries as  summarized in the topological periodic table \cite{SchnyderRFL2008}. % [Table.\ref{table:periodictable}]. 
One may wonder what kind of geometry will describe such TIs. 
%In this presentation, we will discuss the NCG of the A-class TIs based on Ref.\cite{?}.   

Recently two groups, \cite{NeupertSRChMRB2012} and \cite{EstienneRB2012}, independently proposed the quantum Nambu bracket \cite{Nambu1973} for higher dimensional topological insulators. In Ref.\cite{NeupertSRChMRB2012}, the authors considered the quantum Nambu 3-bracket  for chiral TIs (AIII-class). Meanwhile  the authors in Ref.\cite{EstienneRB2012} adopted the the even dimensional quantum Nambu-bracket for A-class TIs, and  3-bracket for 3D TI. 

Inspired by the recent developments, we discuss A-class TIs with emphasis on their relation to higher D. QHE and NCG.  The A-class TIs live in arbitrary even dimensional spaces and do not respect any symmetries such as  time-reversal, particle-hole, and chiral symmetries like QHE.   
Indeed QHE is a 2D entity of A-class TIs, and the A-class TIs can be regarded as higher D counterparts of the QHE, $i.e.$ higher D. QHE. 
The higher D. QHE was first constructed by Zhang and Hu in 4D \cite{ZhangHu2001} as a natural generalization of 2D QHE on Haldane's sphere \cite{Haldane1983}. Their 4D model was soon generalized in even higher dimensions\cite{KarabaliNair2002, HasebeKimura2003}, such as complex projective spaces and even dimensional spheres.  In the modern point of view, the higher D. QHE can be considered as a realization of the A-class TI with Landau levels.  
NCG naturally fits in even dimensional space since each commutator needs a pair of NC coordinates, and so all of the even dimensional coordinates of A-class TIs are neatly fitted in the commutators and NCG is realized in the whole space.   
Thus,  A-class TIs are  a good starting point to see what physical implications  the higher D. NCG brings.  We will discuss the spherical higher D. A-class TIs   
\cite{Hasebe2014-1} on the basis of the former works \cite{HasebeKimura2003}. 
%This presentation is based on the work \cite{Hasebe2014-1}. 

%%%%%%%%%%%%%%%%%%%%%%%%%%%%%%%%%%%%%%%%%%%%%%%%%%%%%%%%%%%%%%%%%%%%%%%%%%%%%%%%%%%%%%%%%%%%
\section{Non-Commutative Geometry of Higher Dimensional Fuzzy Sphere}
%%%%%%%%%%%%%%%%%%%%%%%%%%%%%%%%%%%%%%%%%%%%%%%%%%%%%%%%%%%%%%%%%%%%%%%%%%%%%%%%%%%%%%%%%%%%%

The coordinates of fuzzy two-sphere, $S_F^2$, are defined so as to satisfy the $SU(2)$ algebra \cite{berezin1975}: 
%%%%%%%%%%%%%%%%%%%%%%%%%%%%%%
\be
[X_i, X_j]=i\alpha\epsilon_{ijk}X_k,    \hspace{0.5cm}(i,j,k=1,2,3) 
\label{fuzzytwosphere}
\ee
%%%%%%%%%%%%%%%%%%%%%%%%%%%%%%%%
where $\alpha=2r/I$. ($I/2$ corresponds to the $SU(2)$ spin magnitude.)  
In general, the coordinates of $2k$D fuzzy sphere $S_F^{2k}$  satisfy \cite{Grosse-K-P-1996} 
%%%%%%%%%%%%%%%%%%%%%%%%%%%
\be
[X_a, X_b]=i\alpha X_{ab}, \hspace{0.5cm}(a,b=1,2,\cdots, 2k+1)
\label{fuzzy2ksphere}
\ee
%%%%%%%%%%%%%%%%%%%%%%%%%%
where $X_{ab}$ denote the $SO(2k)$ operators.  For 3D, the 2rank tensor is equivalent to the vector by the relation $X_{ij}=\epsilon_{ijk}X_k$, and then (\ref{fuzzytwosphere}) is realized as a special case of (\ref{fuzzy2ksphere}). Except for the special 3D case $(k=1)$, Eq.(\ref{fuzzy2ksphere}) is not closed only by $X_a$ themselves, but the extra operators $X_{ab}$ are introduced on the right-hand side. There is another mathematical formulation for $S_F^{2k}$ by using the Nambu bracket \cite{Jabbari2004}: 
%%%%%%%%%%%%%%%%%%%%%%%%%%%%%%%%%%%%%
\be
[X_{a_1}, X_{a_2}, \cdots, X_{a_{2k}} ]=i^{k}C(k, I)\alpha^{2k-1}\epsilon_{a_1 a_2 \cdots a_{2k+1}}X_{a_{2k+1}},  
\label{nambu2ksphere}
\ee
%%%%%%%%%%%%%%%%%%%%%%%%%%%%%%%%%%%%%%%
where  $C(k,I)=\frac{(2k)!!(I+2k-2)!!}{2^{2k-1}I!!}$ and $[X_{a_1}, X_{a_2}, \cdots, X_{a_{2k}} ]\equiv X_{[a_1}, X_{a_2}, \cdots, X_{a_{2k}]}$. Eq.(\ref{nambu2ksphere}) is more elegant than Eq.(\ref{fuzzy2ksphere}) in the sense that the $SO(2k)$ operators do not apparently appear the algebra.  Thus, there are two superficially different mathematical formulations for higher D. fuzzy sphere. 

 A simplest way to see the corresponding monopole set-up is to find the right-hand side of the NC algebra. From $S_F^2$ (\ref{fuzzytwosphere}), we can read off the Dirac monopole field strength: 
%%%%%%%%%%%%%%%%%%%%%
\be
F_i\simeq \frac{1}{r^3}x_i. \hspace{0.5cm}(i=1,2,3)
\ee
%%%%%%%%%%%%%%%%%%%%%%%%5
Similarly from (\ref{fuzzy2ksphere}) and (\ref{nambu2ksphere}), we obtain the $SO(2k)$ non-abelian monopole field strength \cite{Yang1978} and the $U(1)$ tensor monopole field strength \cite{Yang1978}: 
%%%%%%%%%%%%%%%%%%%%%%%%%%%%
\bse
\begin{align}
&F_{\mu\nu}\simeq   \frac{1}{r^2}\Sigma_{\mu\nu},  \hspace{0.5cm}(\mu,\nu=1,2,\cdots, 2k) \\
&G_{a_1 a_2 \cdots a_{2k}}\simeq \frac{1}{r^{2k+1}}\epsilon_{a_1 a_2 \cdots a_{2k+1}}x_{a_{2k+1}}, \hspace{0.5cm}(a_1, a_2, \cdots, a_{2k}=1,2,\cdots, 2k+1)
\end{align}
\ese
%%%%%%%%%%%%%%%%%%%%%%%%%%%%%%%%%%
where $\Sigma_{\mu\nu}$ denote the $SO(2k)$ matrices.

%%%%%%%%%%%%%%%%%%%%%%%%%%%%%%%%%%%%%%%%%%%%%%%%%%%%%%%%%%%%
%\begin{figure}[tbph]\center
%\includegraphics*[width=130mm]{Relationsfig.eps}
%\caption{Correspondence between mathematics and physics of  higher dimensional quantum Hall effects and A-class topological insulators. This is taken from Ref.\cite{Hasebe2014-1}.}
%\label{EntirePicture}
%\vspace{-3mm}
%\end{figure}
%%%%%%%%%%%%%%%%%%%%%%%%%%%%%%%%%%%%%%%%%%%%%%%%%%%%%%%%%%%%%% 

%%%%%%%%%%%%%%%%%%%%%%%%%%%%%%%%%%%%%%%%%%%%%%%%%%%%%%%%%%%%%%%%%%%%%%%%%%%%%%%%%%%%%%%%%%%%
\section{Non-Abelian and Tensor Monopole Correspondence}
%%%%%%%%%%%%%%%%%%%%%%%%%%%%%%%%%%%%%%%%%%%%%%%%%%%%%%%%%%%%%%%%%%%%%%%%%%%%%%%%%%%%%%%%%%%%%

Corresponding to the two different NC formulations, we obtain two different monopole set-ups. 
However, since the two NC formulations describe the same NC manifold, $i.e.$ $S_F^{2k}$, the corresponding two monopole set-ups should be same in some sense. Indeed there exists a simple correspondence:
%%%%%%%%%%%%%%%%%%%%%%%%%%%%%%%%%%%%
\be
G_{2k}=\text{tr} (F^{2k}). 
\label{localcorrefandg}
\ee
%%%%%%%%%%%%%%%%%%%%%%%%%%%%%%%%%%%%%
For the fully symmetric representation $[I/2, I/2, \cdots, I/2]$, the right-hand side of (\ref{localcorrefandg}) gives 
%%%%%%%%%%%%%%%%%%%%%%%%%%%%%%%%%%%%%
\be
\text{tr} (F^{2k})=\frac{1}{2^{k+1}r^{2k+1}}c_k(I)\epsilon_{a_1 a_2 \cdots a_{2k+1}}x_{a_{2k+1}}dx_{a_1} dx_{a_2} \cdots dx_{a_{2k}}, 
\label{tracef2k}
\ee
%%%%%%%%%%%%%%%%%%%%%%%%%%%%%%%%%%%%%
where 
%%%%%%%%%%%%%%%%%%%%%%%%%%%%%%%%%%%%%%%%%%%%%
%\be
$c_k(I)=\prod_{l=1}^k \prod_{i=1}^{l}\frac{I+l+i-2}{l+i-1}$ 
%\ee
%%%%%%%%%%%%%%%%%%%%%%%%%%%%%%%%%%%%%%%%%%%%
denotes the $k$th Chern-number of the $SO(2k)$ monopole on $S^{2k}$. Eq.(\ref{tracef2k}) is  actually equivalent to the tensor monopole field strength: 
%%%%%%%%%%%%%%%%%%%%%%%%%%%%%
\be
G_{2k}=g_k\frac{1}{(2k)!r^{2k+1}}\epsilon_{a_1 a_2 \cdots a_{2k+1}}x_{a_{2k+1}}dx_{a_1}dx_{a_2}\cdots dx_{a_{2k}}, 
\ee
%%%%%%%%%%%%%%%%%%%%%%%%%%%%%%%
with tensor monopole charge %$g_k$: 
%%%%%%%%%%%%%%%%%%%%%%%%%%%%
%\be
$g_k=\frac{(2k)!}{2^{k+1}}c_k(I).$ 
%\ee
%%%%%%%%%%%%%%%%%%%%%%%%%%%
%Since the equivalence between their field strengths has been obtained above, 
The monopole gauge fields are similarly related as  
%%%%%%%%%%%%%%%%%%%%%%%%%%%%%%
\be
C_{2k-1}=\text{tr}(L_{\text{CS}}^{(2k-1)}[A]), 
\label{canda2k-1}
\ee
%%%%%%%%%%%%%%%%%%%%%%%%%%%%
where $L_{\text{CS}}^{2k-1}[A]$ denotes the Chern-Simons action in $(2k-1)$D:  
%%%%%%%%%%%%%%%%%%%%%%%%%%%%%
\be
L_{\text{CS}}^{(2k-1)}[A]= k\int_0^1 dt~ \text{tr}(A(tdA+it^2A^2)^{k-1}).  
\label{generalexcernsimonsterm}
\ee
%%%%%%%%%%%%%%%%%%%%%%%%%%%%%%
For instance, $L_{\text{CS}}^{(1)}[A]=\text{tr}A$, $L_{\text{CS}}^{(3)}[A]=\text{tr}( AdA +\frac{2}{3}i A^3)$ and $L_{\text{CS}}^{(5)}[A]
=\text{tr}( A(dA)^2 +\frac{3}{2}iA^3dA-\frac{3}{5} A^5)$. 
By substituting the $SO(2k)$ gauge field to the right-hand side of (\ref{canda2k-1}), we have 
%%%%%%%%%%%%%%%%%%%%%%%%%%%%
\begin{align}
C_1&=-c_1(I)\frac{1}{2r(r+x_3)}\epsilon_{ij3 }x_jdx_i,%~~~~~~~~~~~~~~~~~~~~~~~~~~~~~~~~~~~~~~~~~~~~~~~~~~~~~~~~~~~~~~~~~~~~~~(i=1,2,3)
\nonumber\\
C_{3}&=-c_2(I)\frac{1}{6r^3}\biggl(\frac{1}{r+x_5}+\frac{r}{(r+x_5)^2} \biggr)\epsilon_{abcd 5}x_ddx_a dx_b dx_c, %~~~~~~~~~~~~~~~~~~~~~~~~~~~~~~~~~~~(a,b,c, d=1,2,\cdots,5) 
\nonumber\\
C_{5}&=-c_3(I)\frac{3}{40r^5}\biggl(\frac{1}{r+x_7}+\frac{r}{(r+x_7)^2}+\frac{2}{3}\frac{r^2}{(r+x_7)^3}\biggr)\epsilon_{abcdef7} x_{f}dx_a dx_b dx_c dx_d dx_e. %, %~~~~~~~~~~~~(a,b,\cdots, f=1,2,\cdots,7) 
%\nonumber\\
%C_{a_1 a_2\cdots a_7}
%&=-
%\frac{180}{r^7}\biggl(\frac{1}{r+x_9}
%+\frac{r}{(r+x_9)^2}+\frac{4}{5}\frac{r^2}{(r+x_9)^3}+\frac{2}{5}\frac{r^3}{(r+x_9)^4}\biggr)\epsilon_{a_1a_2\cdots a_8 9 }x_{a_8}. \nonumber% \nn\\
%&~~~~~~~~~~~~~~~~~~~~~~~~~~~~~~~~~~~~~~~~~~~~~~~~~~~~~~~~~~~~~~~~~~~~~~~~~~~~~(a_1,a_2,\cdots, a_7=1,2,\cdots,9)
\label{explicittensorlow}
\end{align}
%%%%%%%%%%%%%%%%%%%%%%%%%%%%%
$C_3$ and $C_5$ represent a natural generalization of the Dirac monopole gauge field $C_1$.  
In general, the $(2k-1)$ rank tensor monopole gauge field exhibits the $k$th power string-like singularity.

%%%%%%%%%%%%%%%%%%%%%%%%%%%%%%%%%%%%%%%%%%%%%%%%%%%%%%%%%%%%
%\begin{figure}[tbph]\center
%\includegraphics*[width=100mm]{SO2kflux.eps}
%\caption{The internal geometry the $SO(2k)$ non-abelian flux is equivalent to the fuzzy-fibre  $S_F^{2k-2}$, and the $S_F^{2k-2}$  corresponds to $(2k-2)$-brane in the enlarged $(4k-2)$ dimensional  space. This is taken from Ref.\cite{Hasebe2014-1}. }
%\label{so2kflux}
%\vspace{-3mm}
%\end{figure}
%%%%%%%%%%%%%%%%%%%%%%%%%%%%%%%%%%%%%%%%%%%%%%%%%%%%%%%%%%%%%%

%%%%%%%%%%%%%%%%%%%%%%%%%%%%%%%%%%%%%%%%%%%%%
%\subsection{Non-commutative Geometry}
%%%%%%%%%%%%%%%%%%%%%%%%%%%%%%%%%%%%%%%%%%%%%

In the lowest Landau level (LLL), the covariant angular momentum vanishes and so the total angular momentum is identified with the field strength: 
%%%%%%%%%%%%%%%%%%%%%%%%%%
\be
L_{ab}=\Lambda_{ab}+r^2F_{ab}~~\sim ~~r^2 F_{ab}. 
\label{labfabident}
\ee
%%%%%%%%%%%%%%%%%%%%%%%%%%%
According to (\ref{localcorrefandg}), we have  
%%%%%%%%%%%%%%%%%%%%%%%%%%%%%
%\be
$\frac{1}{r^{2k+1}} x_a =\frac{2}{(2k)!c_k(I)}\epsilon_{a a_1 a_2 \cdots a_{2k}}\text{tr} (F_{a_1 a_2 } \cdots F_{a_{2k-1} a_{2k}}),$    
%\label{xfromfs}
%\ee
%%%%%%%%%%%%%%%%%%%%%%%%%%%
and with (\ref{labfabident}), we find that the coordinates in the LLL become to  
%%%%%%%%%%%%%%%%%%
\be
X_a=\frac{I}{(2k)! c_k(I)} ~{\alpha}~\epsilon_{a a_1 a_2 \cdots a_{2k}}(L_{a_1 a_2}L_{a_3 a_4} \cdots L_{a_{2k-1} a_{2k}}),   
\label{xafromlas}
\ee
%%%%%%%%%%%%%%%%%%%
which obey the quantum Nambu algebra (\ref{nambu2ksphere}). Thus, the quantum Nambu geometry naturally emergies in the LLL of  A-class TIs. 

%%%%%%%%%%%%%%%%%%%%%%%%%%%%%%%%%%%%%%%%%%%%%%%%%%%%%%%%%%%%%%%%%%%%%%%%%%%%%%%%%%%%%%%%%%%%
\section{Tensor Chern-Simons Theory for A-Class Topological Insulator}
%%%%%%%%%%%%%%%%%%%%%%%%%%%%%%%%%%%%%%%%%%%%%%%%%%%%%%%%%%%%%%%%%%%%%%%%%%%%%%%%%%%%%%%%%%%%%

The non-Abelian and tensor monopole correspondence brings important physical implications to the A-class topological insulators. 
In the original QHE, the Chern-Simons (CS) flux attachment is introduced to cancel the external $U(1)$ magnetic field: 
%%%%%%%%%%%%%%%%%%%%%%%%%%%%
\be
A_1-C_1=0. 
\ee
%%%%%%%%%%%%%%%%%%%%%%%%%%%%% 
This cancellation is rather trivial in the sense both external and Chern-Simons gauge fields are Abelian. 
In higher dimensions, the external magnetic fields are non-Abelian gauge fields. The non-Abelian and tensor monopole correspondence tells that the non-abelian magnetic field can be canceled by  $\it{Abelian}$ gauge field:  
%%%%%%%%%%%%%%%%%%%%%%%%%%%%%%%%%%%%%%
\be
\text{tr}(L_{\text{CS}}^{(2k-1)}[A])-C_{2k-1}=0. 
\ee
%%%%%%%%%%%%%%%%%%%%%%%%%%%%%%%%%%%%%%%
For instance $k=2,3$, we have 
%%%%%%%%%%%%%%%%%%%%%%%%%%%%%%%%%%%%
\begin{align}
&\text{tr}( AdA +\frac{2}{3}i A^3)-C_3=0, \nonumber\\
&\text{tr}( A(dA)^2 +\frac{3}{2}iA^3dA-\frac{3}{5} A^5)-C_5=0. 
\end{align}
%%%%%%%%%%%%%%%%%%%%%%%%%%%%%%%%%%%%%
Due to this notable cancellation, we can generalize exotic ideas cultivated in the QHE to A-class TIs, such as topological field theory description of QHE \cite{Zhang-H-K-1988}.   
We propose the following tensor-type Chern-Simons field theory for the A-class TIs: 
%. Flux attachment to membrane is readily accomplished by introducing the tensor-type Chern-Simons field theory: 
%%%%%%%%%%%%%%%%%%%%%%%%%%%%%%%%%%%%%%
\be
S=\int_{4k-1} C_{2k-1}J_{2k}+\frac{1}{4\pi m^k}\int_{4k-1}C_{2k-1}G_{2k},  
\label{tensorChernSimonsaction}
\ee
%%%%%%%%%%%%%%%%%%%%%%%%%%%%%%%%%%%%%%
where $G_{2k}=dC_{2k-1}$ and $m$ denotes an odd integer in accordance with the filling factor 
%%%%%%%%%%%%%%%%%%%%
%\be
$\nu_{2k}={1}/{m^k}.$  
%\ee
%%%%%%%%%%%%%%%%%%%%%
The membranes described by the tensor Chern-Simons action (\ref{tensorChernSimonsaction}) generally obey the fractional statistics related the linking number \cite{NepomechieZee1984}.   
The equations of motion of the Chern-Simons field yield 
%%%%%%%%%%%%%%%%%%%%%%
\be
J_{2k}=-\frac{1}{2\pi m^k} G_{2k}. 
\label{equationforcp+1}
\ee
%%%%%%%%%%%%%%%%%%%%%%%
For component representation, Eq.(\ref{equationforcp+1}) can be rewritten as 
%%%%%%%%%%%%%%%%%%%%%%%%%%%%%%%
%\be
\begin{subequations}
\begin{align}
&
%\hspace{-3mm}
J^{i_1i_2\cdots i_{2k-2} 0} =-\frac{1}{2\pi m^k}B^{i_i i_2\cdots i_{2k-2}}, \\%\hspace{5mm}
&
J^{i_1i_2\cdots i_{2k-1}} =\frac{1}{ 2\pi (2k)! m^k}\epsilon^{i_1i_2\cdots i_{4k-2}}E_{i_{2k}\cdots i_{4k-2}}, %\label{halleffectgene2}
\end{align}
\label{generalhallflux}
%\ee
\end{subequations}
%%%%%%%%%%%%%%%%%%%%%%%%%%%%%%%
Eqs.(\ref{generalhallflux})  represent  generalized flux attachment and Hall effect.  
The $(2k-1)$-form Chern-Simons field is naturally coupled to  $(2k-2)$ dimensional membrane-like object, $(2k-2)$-brane.   Since $m^k$ is an odd integer, the odd number fluxes are attached to membrane and the membrane becomes a composite object of the original membrane and the fluxes. Since at $\nu_{2k}=1$, membrane corresponds to a fermion,  the  odd number  ${m^k}$ flux attachment transmutes the statistics of the membrane to Bose statistics just like composite boson.    
The groundstate of the A-class TI can be regarded as a superfluid state of the composite membranes. 
The fractionally charged quasi-particle in fractional QHE \cite{ArovasSW1984} is also  generalized to   $(2k-2)$-brane excitation with a fractional charge: 
%%%%%%%%%%%%%%%%%%%%%%%%%
\be
e_{2k-2}^*={1}/{m^k}.  
\ee
%%%%%%%%%%%%%%%%%%%%%%%%%
It turns out that the fractional charged membrane obeys the fractional statistics mediated by the tensor gauge field. 
Particularly interesting property is the dimensional hierarchy.  
From the 0-brane picture, the total filling factor is expressed as 
%%%%%%%%%%%%%%%%%%%%%%%%%%%
\be
\nu=\nu_{2}\nu_4\nu_6 \cdots \nu_{2k}={1}/{m^{\frac{1}{2}k(k+1)}}. 
\label{dimhiernu}
\ee
%%%%%%%%%%%%%%%%%%%%%%%%%%%
The hierarchical structure of the filling factor may remind the Haldane-Halperin hierarchy of the fractional QHE \cite{Haldane1983, Halperin1984}, however notice that the hierarchy (\ref{dimhiernu}) ranges over different dimensions not in the same dimension.   According to the Haldane-Halperin's idea of the hierarchy, the hierarchical structure of the filling fraction  implies the condensation of quasi-excitations to form a new generation of incompressible liquid. The dimensional hierarchy of the filling fraction implies that  the low dimensional membrane-like objects condense to make a higher dimensional quantum incompressible liquid of A-class TIs.

%%%%%%%%%%%%%%%%%%%%%%%%%%%%%%%%%%%%%%%%%%%%%%%%%%%%%%%%%%%%%%%%%%%%%%%%%%%%%%%%%%%%%%%%%%%%
\section{Summary}
%%%%%%%%%%%%%%%%%%%%%%%%%%%%%%%%%%%%%%%%%%%%%%%%%%%%%%%%%%%%%%%%%%%%%%%%%%%%%%%%%%%%%%%%%%%%%

\begin{figure}[tbph]\center %\hspace{0.1cm}
\includegraphics*[width=120mm]{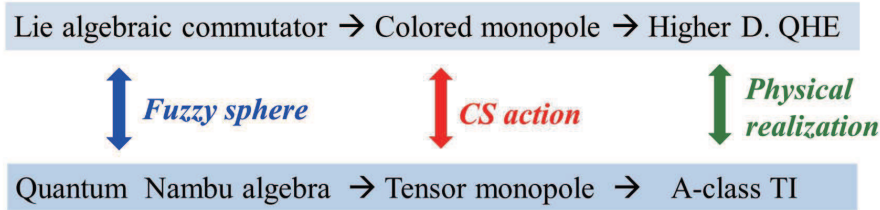}
\caption{The fuzzy sphere has two superficially different mathematical formulations. Corresponding to the two realizations, there are two monopole set-ups related by the Chern-Simons action. Higher D. QHE is a realization of the A-class TI with Landau levels.  }
\label{summaryobservation}
%\vspace{-10mm}
\end{figure}

 We exploited the A-class TI with emphasis on its relations to quantum Nambu geometry. 
 The main observations are summarized in Fig.\ref{summaryobservation}. We showed  
 that the exotic physics of QHE can naturally be generalized in A-class TIs based on the non-Abelian and tensor monopole correspondence.  
%Interested readers may consult Ref.\cite{Hasebe2014-1} for more details.  
 While we focused on the A-class TIs, similar discussions can be expanded for AIII-class TIs that live in arbitrary odd dimensional spaces \cite{Hasebe2014-2}.

%%%%%%%%%%%%%%%%%%%%%%%%%%%%%%%%%%%%%%%%%%%%%%%%%%%%%%%%%%%%
%%%%%%%%%%%%%%%%%%%%%%%%%%%%%%%%%%%%%%%%%%%%%%%%%%%%%%%%%%%%%%

%%%%%%%%%%%%%%%%%%%%%%%%%%%%%%%%%%%%%%%%%%%%%%%%%%%%%%%%%%%%%%%%%%%%%%%%%%
\begin{acknowledgement}
%%%%%%%%%%%%%%%%%%%%%%%%%%%%%%%%%%%%%%%%%%%%%%%%%%%%%%%%%%%%%%%%%%%%%%%%%%%%
  
This work was partially supported by a Grant-in-Aid for Scientific Research  from the Ministry of Education, Science, Sports and Culture of Japan (Grant No.23740212), Overseas   Dispatching Program 2013 of National College of Technology, and The Emma Project for Art and Culture.

%%%%%%%%%%%%%%%%%%%%%%%%%%%%%%%%%%%%%%%%%%%%%%%%%%%%%%%%%%%%%%%%%%%%%%%%%%%%
\end{acknowledgement}
%%%%%%%%%%%%%%%%%%%%%%%%%%%%%%%%%%%%%%%%%%%%%%%%%%%%%%%%%%%%%%%%%%%%%%%%%%%%%

%The style of the following references should be used in all documents.

%%%%%%%%%%%%%%%%%%%%%%%%%%%%%%%%%%%%%%%%%%%%%%%%%%%%%%%%%%%%%%%%%%%%%%%%%

\end{document}